\documentclass{jpsj3}
\usepackage{txfonts}
\usepackage{graphicx}
\usepackage{dcolumn}
\usepackage{bm}

\font\scripti=cmmi7
\font\scriptscripti=cmmi5
\def\sib#1{\setbox0 = \hbox{\scripti #1}
  \kern-.02em\copy0\kern-\wd0
  \kern.04em\box0} 
\def\ssib#1{\setbox0 = \hbox{\scriptscripti #1}
  \kern-.02em\copy0\kern-\wd0
  \kern.04em\box0} 
\font\tenib=cmmib10 
\skewchar\tenib='177 \skewchar\tenib='177 \skewchar\tenib='177
\def\pbold#1{\setbox0 = \hbox{$ #1 $}
  \kern-.022em\copy0\kern-\wd0
  \kern.011em\copy0\kern-\wd0
  \kern.011em\copy0\kern-\wd0
  \kern.011em\copy0\kern-\wd0
  \kern.011em\box0} 

\def\lesssim{\ \raise.3ex\hbox{$<$}\kern-0.8em\lower.7ex\hbox{$\sim$}\ }
\def\gesim{\ \raise.3ex\hbox{$>$}\kern-0.8em\lower.7ex\hbox{$\sim$}\ }
\title{Single-particle Excitations and Effects of Hetero-pairing Fluctuations in a Bose-Fermi Mixture with a Feshbach Resonance}

\author{Digvijay Kharga\thanks{digvijay@rk.phys.keio.ac.jp}, Daisuke Inotani, Ryo Hanai, Yoji Ohashi}
\inst{Department of Physics, Keio University, 3-14-1 Hiyoshi, Kohoku-ku, Yokohama 223-8522, Japan} 

\abst{We theoretically investigate normal-state properties of a gas mixture of single-component bosons and fermions with a hetero-nuclear Feshbach resonance. Including strong hetero-pairing fluctuations associated with the Feshbach resonance, we calculate single-particle density of states, as well as the spectral weight at various interaction strengths. For this purpose, we employ an improved $T$-matrix approximation (TMA), where the bare Bose Green's function in the non-selfconsistent TMA self-energy is modified so as to satisfy the Hugenholtz-Pines relation at the Bose-Einstein condensation (BEC) temperature $T_{\rm BEC}$. In the unitary regime at $T_{\rm BEC}$, we show that hetero-pairing fluctuations couple Fermi atomic excitations with Fermi molecular excitations, as well as with Bose atomic excitations. Although a similar coupling phenomenon by pairing fluctuations is known to give a pseudo-gapped density of states in the unitary regime of a two-component Fermi gas, such a dip structure is found to not appear even in the unitary limit of a Bose-Fermi mixture. It only appears in the strong-coupling regime. Instead, a spectral peak along the molecular dispersion appears in the spectral weight. We also clarify how this coupling phenomenon is seen in the Bose channel. Since a hetero-nuclear Feshbach resonance, as well as the formation of Bose-Fermi molecules, have been realized, our results would be useful for the study of strong-coupling properties of this unique quantum gas.} 
\begin{document}
\maketitle
\par
\section{Introduction}
\par
In cold atom physics, a gas mixture of single-component bosons and fermions\cite{Inguscio,Modugno,Stan,Inouye,Zaccanti,Ferlaino,Bongs,Deh,Schuster,Repp} has attracted much attention as a counterpart of two-component Fermi gas\cite{Regal2004,Zwierlein2004,Bartenstein2004,Kinast2004,Levin,Bloch,Giorgini}. In both the cases, one can tune the strength of a pairing interaction between different species by using a Feshbach resonance\cite{Pethick,Chin}. In the Fermi-Fermi case, this unique technique has extensively been used to study BCS (Bardeen-Cooper-Schrieffer)-BEC (Bose-Einstein condensation) crossover physics\cite{Regal2004,Zwierlein2004,Bartenstein2004,Kinast2004}, where the character of Fermi superfluid continuously changes from the weak-coupling BCS-type to BEC of tightly bound molecules, with increasing the interaction strength\cite{Eagles,Leggett,NSR,Melo,Randeria,Engelbrecht,Strinati,Ohashi,Perali}. In the intermediate coupling regime (BCS-BEC crossover region), strong pairing fluctuations are expected to cause the so-called pseudogap phenomenon\cite{Tsuchiya2009,Watanabe2010,Hui2010,Chien2010,Perali2011,Chen2014,Miki}. Although this expectation has not completely been confirmed experimentally yet, the recent photoemission-type experiments on $^{40}$K Fermi gases have observed a back-bending behavior of single-particle excitations\cite{Stewart2008,Gaebler2010,Sagi2015}, being consistent with the pseudogap scenario\cite{Tsuchiya2009,Watanabe2010,Hui2010,Chien2010,Perali2011,Chen2014,Miki}. Since the pseudogap in this case originates from the formation of (fluctuating) preformed Cooper pairs, it is interesting to explore a similar phenomenon caused by hetero-pairing fluctuations in a Bose-Fermi mixture. In addition, while preformed Cooper pairs (or pairing fluctuations) in a two-component Fermi gas are bosonic, hetero-pairing fluctuations in a Bose-Fermi mixture are fermionic, in the sense that they eventually change to molecular fermions in the strong-coupling limit. Thus, it is also an interesting problem how this quantum-statistical difference is reflected in strong-coupling properties of a Bose-Fermi mixture.
\par
The purpose of this paper is to theoretically investigate single-particle properties of a Bose-Fermi mixture with a hetero-nuclear Feshbach resonance. Including Bose-Fermi hetero-pairing fluctuations tuned by a Feshbach resonance, we calculate the single-particle density of state, as well as the single-particle spectral weight, in both the Bose and Fermi channels. We clarify strong-coupling corrections to these quantities, from the weak-coupling regime to the strong-coupling regime in the normal state above the Bose-Einstein condensation temperature $T_{\rm BEC}$.
\par
In cold Fermi gas physics, a (non self-consistent) $T$-matrix approximation (TMA) has frequently been used to deal with strong pairing fluctuations in the BCS-BEC crossover region\cite{Tsuchiya2009,Watanabe2010,Hui2010,Perali2011,Miki}. The pseudogap phenomenon in this system has been predicted by using this strong-coupling theory\cite{Tsuchiya2009,Watanabe2010}. TMA has also succeeded in explaining the photoemission spectra\cite{Miki,Tsuchiya2011} observed in $^{40}$K Fermi gases\cite{Stewart2008,Gaebler2010,Sagi2015}. 
\par
\begin{figure}
\begin{center}
\includegraphics[width=9cm]{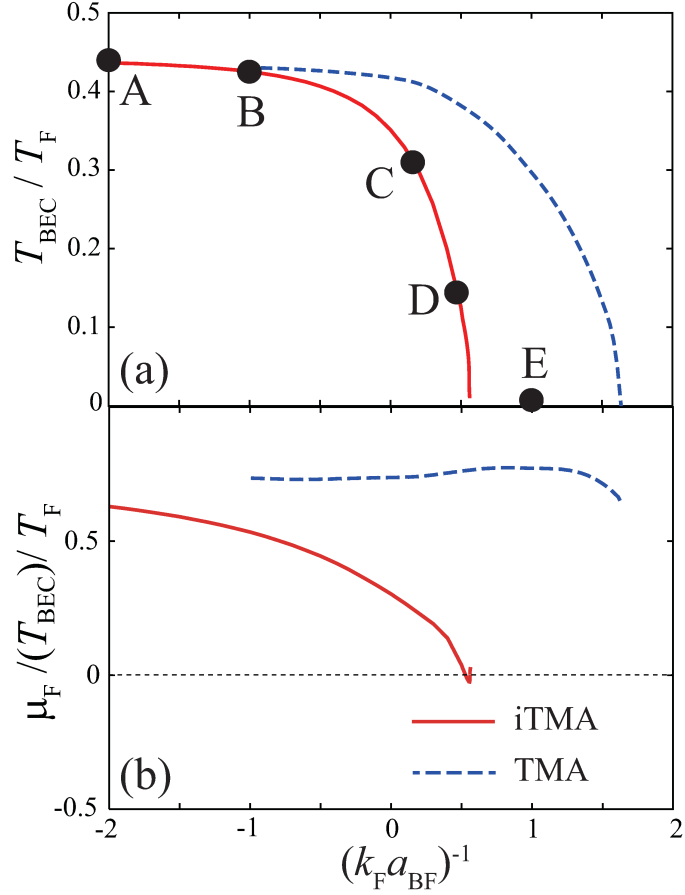}
\caption{(Color online) Calculated BEC phase transition temperature $T_{\rm BEC}$ (a), as well as the Fermi chemical potential $\mu_{\rm F}(T_{\rm BEC})$ (b), in a Bose-Fermi mixture. iTMA: improved TMA developed by the authors\cite{Vijay}. TMA: ordinary non self-consistent $T$-matrix approximation\cite{Ref3}. The strength of an inter-species interaction is measured in terms of the inverse $s$-wave scattering length $a_{\rm BF}^{-1}$, normalized by the Fermi wavelength $k_{\rm F}$. $T_{\rm F}$ is the Fermi temperature. In panel (a), A$\sim$E show the interaction strengths at which we examine single-particle excitations in this paper. In iTMA, $T_{\rm BEC}$ vanishes, when $(k_{\rm F}a_{\rm BF})^{-1}\simeq 0.56$.
}
\label{fig1}
\end{center}
\end{figure}
\par
However, we have recently pointed out that TMA has room for improvement\cite{Vijay}, when it is applied to a Bose-Fermi mixture. To explain this, we first recall that the superfluid phase transition temperature $T_{\rm c}$ in an ultracold Fermi gas can conveniently be determined from the Thouless criterion\cite{Thouless}, stating that the particle-particle scattering matrix $\Gamma_{\rm FF}({\bm q},\omega)$ diverges in the low-energy and long wavelength limit ($\omega={\bm q}=0$) at $T_{\rm c}$. Because this scattering matrix $\Gamma_{\rm FF}({\bm q},\omega)$ is involved in the TMA self-energy\cite{Perali}, strong pairing fluctuations near $T_{\rm c}$, as well as their effects on single-particle excitations, are treated in a consistent manner, when TMA is applied to an ultracold Fermi gas. 
\par
On the other hand, in the case of Bose-Fermi mixture, $T_{\rm BEC}$ is determined from the Hugenholtz-Pines condition\cite{HP}, stating that the {\it dressed} Bose Green's function in TMA has gapless excitations in the BEC phase. However, in the non-selfconsistent TMA, the bare Bose Green's function is used in evaluating the TMA self-energy. As a result, because the bare Bose Green's function still has {\it gapped} excitations even at $T_{\rm BEC}$, strong-coupling effects on the TMA self-energy is underestimated. Indeed, Fig. \ref{fig1} shows that, even in the strong-coupling regime ($(k_{\rm F}a_{\rm BF})^{-1}\sim 1.5>0$, where $a_{\rm BF}$ is an $s$-wave scattering length for an inter-species interaction, and $k_{\rm F}$ is the Fermi wavelength of Fermi atoms) where Bose and Fermi atoms are expected to form two-body bound molecules, the TMA Fermi chemical potential $\mu_{\rm F}(T_{\rm BEC})$ is still {\it positive}\cite{Ref3}, which looks as if unpaired Fermi {\it atoms} still exist, forming a large Fermi surface. 
\par
To cure this, we have recently proposed to replace the bare Bose Green's function in the TMA self-energy by a modified one which satisfies the required Hugenholtz-Pines condition\cite{Vijay}. This replacement naturally enhances low-energy Bose atomic excitations, leading to the remarkable decrease of the Fermi chemical potential $\mu_{\rm F}(T_{\rm BEC})$ around $(k_{\rm F}a_{\rm BF})^{-1}=0.5$\cite{note1}, as shown in Fig. \ref{fig1}(b). In this paper, we employ this improved TMA (iTMA).
\par
For the current stage of research for Bose-Fermi mixtures, hetero-nuclear Feshabch resonances have been observed in various gases, such as $^6$Li-$^7$Li\cite{Salomon}, $^6$Li-$^{87}$Rb\cite{Deh}, $^6$Li-$^{23}$Na\cite{Stan,Schuster}, $^6$Li-$^{133}$Cs\cite{Repp}, and $^{40}$K-$^{87}$Rb\cite{Zaccanti,Ferlaino,Inouye}. Hetero-nuclear molecules have also been produced in a $^{40}$K-$^{87}$Rb mixture loaded on an optical lattice\cite{Bongs2}. A more complicated Bose-Fermi mixture, consisting of two-component fermions and single-component bosons has also been realized, where Bose and Fermi double-condensate has been observed\cite{FFB1,FFB2,Horikoshi} Theoretically, the {\it ordinary} TMA  has been used to deal with hetero-pairing fluctuations\cite{Ref3,Yabu1,Suzuki1,Ref4,Ref5}, to obtain $T_{\rm BEC}$\cite{Ref3,Ref4}, as well as single-particle excitation spectra in the strong-coupling regime at $T=0$\cite{Ref5}. On the viewpoint of chemical equilibrium among Bose atoms, Fermi atoms, and quasi-molecular fermions, Refs.\cite{Yabu2,Yabu3} discuss the phase diagram of a Bose-Fermi mixture. Besides pairing physics, the stability of a Bose-Fermi mixture has also been examined, both theoretically\cite{Molmer,Miyakawa,Roth} and experimentally\cite{Inguscio}.
\par
This paper is organized as follows. In Sec. 2, we explain our formulation (iTMA)\cite{Vijay}. In Sec. 3, we show our results on the single-particle spectral weight, as well as the single-particle density of states in the normal state above $T_{\rm BEC}$, to discuss strong-coupling properties of a Bose-Fermi mixture. Relation to the case of a two-component Fermi gas in the BCS-BEC crossover region is also discussed. Throughout this paper, we set $\hbar=k_{\rm B}=1$, and the system volume $V$ is taken to be unity, for simplicity.
\par
\par
\section{Formulation}
\par
We consider a Bose-Fermi mixture consisting of single-component fermions and bosons, described by the Hamiltonian\cite{Ref3,Vijay},
\begin{eqnarray} 
H&=&\sum_{{\bm p},{\rm s=F,B}}
\xi_{\bm p}^{\rm s}
c_{{\rm s},{\bm p}}^\dagger
c_{{\rm s},{\bm p}}
\nonumber
\\
&-&
U_{\rm BF}
\sum_{{\bm p},{\bm p}',{\bm q}}
c_{{\rm B},{\bm p}+{\bm q}/2}^\dagger
c_{{\rm F},-{\bm p}+{\bm q}/2}^\dagger
c_{{\rm F},-{\bm p}'+{\bm q}/2}
c_{{\rm B},{\bm p}'+{\bm q}/2}.
\label{eq.1}
\end{eqnarray}
Here, $c_{{\rm F},{\bm p}}^{\dagger}$ ($c_{{\rm B},{\bm p}}^{\dagger}$) is the creation operator of a Fermi (Bose) atom, with the kinetic energy $\xi_{\bm p}^{\rm F}={\bm p}^2/(2m_{\rm F})-\mu_{\rm F}$ ($\xi_{\bm p}^{\rm B}=p^2/(2m_{\rm B})-\mu_{\rm B}$), measured from the Fermi (Bose) chemical potential $\mu_{\rm F}$ ($\mu_\text{B}$), where $m_{\rm F}$ ($m_{\rm B}$) is a Fermi (Bose) atomic mass. $-U_{\rm BF}$ ($<0$) is an attractive inter-species interaction, which is assumed to be tunable by adjusting the threshold energy of a hetero-nuclear Feshbach resonance\cite{Chin}. 
\par
In this paper, we ignore effects of a harmonic trap, for simplicity. In addition, although the value of fermion mass $m_{\rm F}$ is different from that of boson mass $m_{\rm B}$ in a real Bose-Fermi mixture, we also ignore this difference, to simply take $m_{\rm F}=m_{\rm B}\equiv m$. (For effects of the mass difference, see, for example, Ref.\cite{Ref4}.) Furthermore, we focus on hetero-pairing fluctuations associated with the inter-species interaction $U_{\rm BF}$, ignoring other intra-species ones. We briefly note that the latter interactions is important in considering the stability of the system\cite{Pethick,Molmer,Miyakawa,Roth}. For the number of atoms, we consider the simplest balanced case, that is, the number $N_{\rm F}$ of Fermi atoms equals the number $N_{\rm B}$ of Bose atoms ($N_{\rm F}=N_{\rm B}\equiv N$). 
\par
As usual, we eliminate the ultraviolet divergence involved in the model Hamiltonian in Eq. (\ref{eq.1}), by measuring the interaction strength in terms of the inverse $s$-wave scattering $a_{\rm BF}^{-1}$, normalized by the Fermi momentum $k_{\rm F}$, as $(k_{\rm F}a_{\rm BF})^{-1}$. The observable scattering length $a_{\rm BF}$ is related to the bare interaction $U_{\rm BF}$ as,
\begin{equation}
{4\pi a_{\rm BF} \over m}=
-{U_{\rm BF} \over 1-U_{\rm BF}\sum_{\bm p}^{p_{\rm c}}{1 \over 2\varepsilon_{\bm p}}},
\label{eq.2}
\end{equation}
where $\varepsilon_{\bm p}={\bm p}^2/(2m)$, and $p_{\rm c}$ is a high-momentum cutoff. In the two-particle case ($N_{\rm F}=N_{\rm B}=1$), a Fermi atom and a Bose atom form a two-body bound state with the binding energy $E_{\rm bind}=-1/(ma_{\rm BF}^2)$, when $(k_{\rm F}a_{\rm BF})^{-1}>0$. Thus, one may physically regard the region $(k_{\rm F}a_{\rm BF})^{-1}>0$~($<0$) as the strong-coupling (weak-coupling) side, although there is actually no clear ``phase boundary" at the unitarity $(k_{\rm F}a_{\rm BF})^{-1}=0$.
\par
\begin{figure}
\begin{center}
\includegraphics[width=12cm]{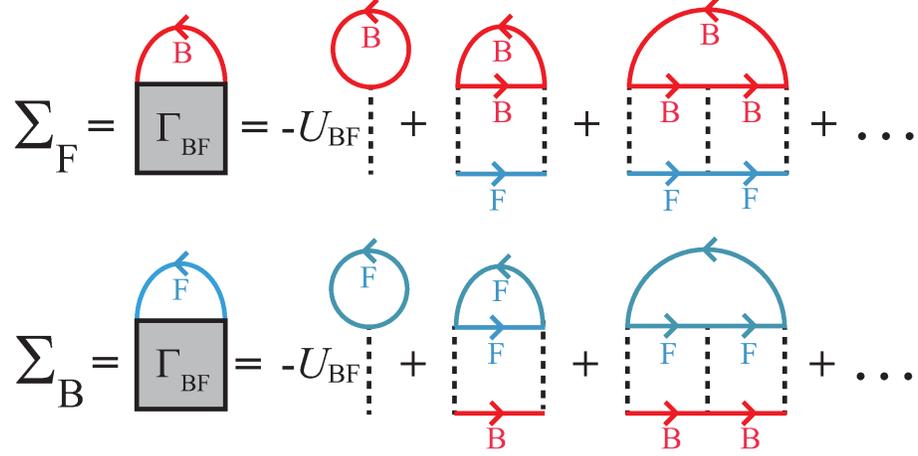}
\caption{(Color online) Self-energies $\Sigma_{{\rm s}={\rm B,F}}({\bm p},i\omega_n^{\rm s})$ in the ordinary (non self-consistent) $T$-matrix approximation (TMA). The solid lines with the label ``B" and ``F" describe the bare Bose Green's function $G_{\rm B}^0$ and the bare Fermi Green's function $G_{\rm F}^0$, respectively. The dashed line is the inter-species interaction $-U_{\rm BF}~(<0)$. $\Gamma_{\rm BF}$ is the TMA boson-fermion scattering matrix. We note that iTMA is achieved by replacing $G_{\rm B}^0$ in the TMA self-energy diagrams with ${\tilde G}_{\rm B}$ in Eq. (\ref{eq.10}).}
\label{fig2}
\end{center}
\end{figure}
\par
Strong-coupling corrections to single-particle excitations can conveniently be described by the self-energies $\Sigma_{{\rm s}={\rm B,F}}({\bm p},i\omega_n^{\rm s})$ in the single-particle Bose (${\rm s}={\rm B}$) and Fermi (${\rm s}={\rm F}$) thermal Green's functions\cite{Fetter},
\begin{equation}
G_{\rm s}({\bm p},i\omega_n^{\rm s}) =
{1 \over 
i\omega_n^{\rm s}-\xi_{\bm p}^{\rm s}
-\Sigma_{\rm s}({\bm p},i\omega_n^{\rm s})}.
\label{eq.3}
\end{equation}
Here, $\omega_n^{\rm F}$ and $\omega_n^{\rm B}$ represent the fermion and boson Matsubara frequencies, respectively. As mentioned previously, this paper employs the improved $T$-matrix approximation (iTMA) developed in Ref.\cite{Vijay}, to evaluate $\Sigma_{{\rm s}={\rm B,F}}({\bm p},i\omega_n^{\rm s})$. To explain this approach, we first briefly review the ordinary non self-consistent TMA, where the self-energies are diagrammatically given as Fig. \ref{fig2}. Summing up these diagrams, we have\cite{Ref3},
\begin{equation}
\Sigma_{\rm B}({\bm p}, i\omega_n^{\rm B})=
T\sum_{{\bm q},\omega_{n'}^{\rm F}}
\Gamma_{\rm BF}({\bm q},i\omega_{n'}^{\rm F})
G_{\rm F}^0({\bm q}-{\bm p},i\omega_{n'}^{\rm F}-i\omega_{n}^{\rm B}),
\label{eq.4}
\end{equation}
\begin{equation}
\Sigma_{\rm F}({\bm p}, i\omega_n^{\rm F})=
-T\sum_{{\bm q},\omega_{n'}^{\rm F}}
\Gamma_{\rm BF}({\bm q},i\omega_n^{\rm F})
G_{\rm B}^0({\bm q}-{\bm p},i\omega_{n'}^{\rm F}-i\omega_{n}^{\rm F}).
\label{eq.5}
\end{equation}
In Eqs. (\ref{eq.4}) and (\ref{eq.5}), hetero-pairing fluctuations are described by the TMA boson-fermion scattering matrix,
\begin{eqnarray}
\Gamma_{\rm BF}({\bm q},i\omega_n^{\rm F})
&=&
-{U_{\rm BF} \over 1-U_{\rm BF}\Pi_{\rm BF}({\bm q},i\omega_n^{\rm F})}
\nonumber
\\
&=&
{
{4\pi a_{\rm BF} \over m}
\over
1+
{4\pi a_{\rm BF} \over m}
\left[
\Pi_{\rm BF}({\bm q},i\omega_n^{\rm F})
-\sum_{\bm p}{1 \over 2\varepsilon_{\bm p}}
\right]
},
\label{eq.6}
\end{eqnarray}
where 
\begin{equation}
\Pi_{\rm BF}({\bm q},i\omega_n^{\rm F})
=T\sum_{{\bm k},\omega_{n'}^{\rm B}}
G_{\rm F}^0({\bm q}-{\bm k},i\omega_n^{\rm F}-i\omega_{n'}^{\rm B})
G_{\rm B}^0({\bm k},i\omega_{n'}^{\rm B}),
\label{eq.7}
\end{equation}
is the lowest-order hetero-pair correlation function. In the second line in Eq. (\ref{eq.6}), the ultraviolet divergence coming from the momentum summation in Eq. (\ref{eq.7}) has been absorbed into the scattering length $a_{\rm BF}$. The fact that the fermion Matsubara frequency appears in $\Gamma_{\rm BF}({\bm q},i\omega_n^{\rm F})$ in Eq. (\ref{eq.6}) reflects the fermionic character of hetero-pairing fluctuations. 
\par
A crucial point in the ordinary TMA is that the bare Bose and Fermi Green's functions,
\begin{equation}
G_{{\rm s}={\rm B,F}}^0({\bm p},i\omega_n^{\rm s})=
{1 \over i\omega_n^{\rm s}-(\varepsilon_{\bm p}-\mu_{\rm s})},
\label{eq.8}
\end{equation}
are used in in Eqs. (\ref{eq.4}), (\ref{eq.5}), and (\ref{eq.7}). As a result, when one determines $T_{\rm BEC}$ from the Hugenholtz-Pines condition\cite{HP},
\begin{equation}
\mu_{\rm B}=\Sigma_{\rm B}({\bm q}=0, i\omega_n^{\rm B}=0),
\label{eq.9}
\end{equation}
although the TMA Bose Green's function $G_{\rm B}$ in Eq. (\ref{eq.3}) has the required gapless excitations at $T_{\rm BEC}$, the bare Bose Green's function $G_{\rm B}^0$ in Eq. (\ref{eq.8}) does not (with the energy gap, $E_{\rm gap}=-\Sigma_{\rm B}(0,0)~(>0)$), so that low-energy Bose-atomic excitations are underestimated in the TMA self-energies $\Sigma_{{\rm s}={\rm B,F}}({\bm q},i\omega_n^{\rm s})$.
\par
\begin{figure}
\begin{center}
\includegraphics[width=9cm]{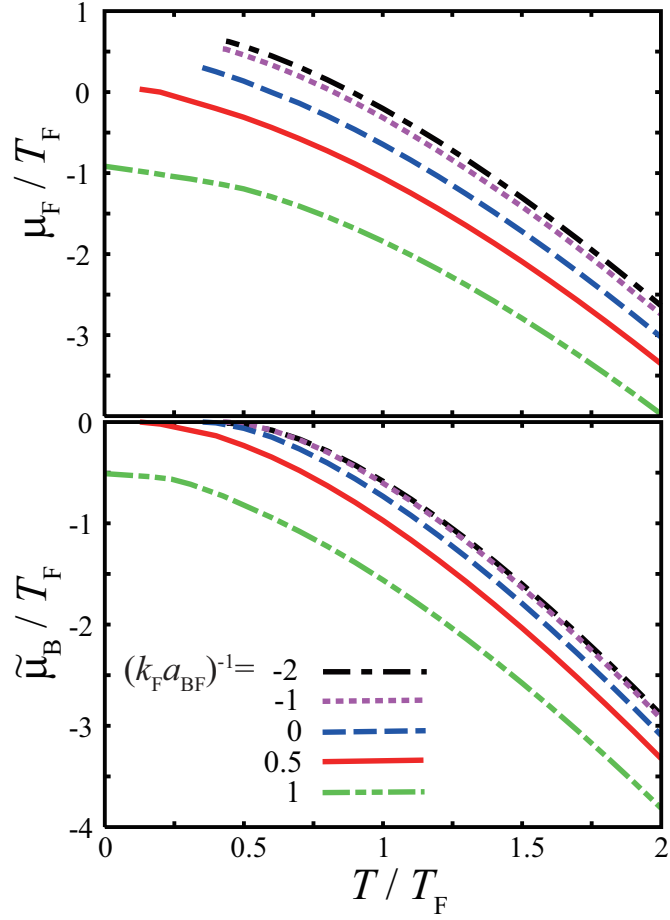}
\caption{(Color online) Calculated Fermi chemical potential $\mu_{\rm F}(T)$ (a), and the effective Bose chemical potential ${\tilde \mu}_{\rm B}\equiv\mu_{\rm B}-\Sigma_{\rm B}(0,0)$ (b), in the normal state of a Bose-Fermi mixture. The interaction strengths used here correspond to A$\sim$E in Fig. \ref{fig1}(a). These data are used in evaluating the single-particle spectral weight $A_{{\rm s}={\rm B,F}}({\bm p},\omega)$, as well as the single-particle density of states $\rho_{{\rm s}={\rm B,F}}(\omega)$ in Sec. 3.}
\label{fig3}
\end{center}
\end{figure}
\par
In the improved TMA (iTMA)\cite{Vijay}, the bare Bose Green's function $G_{\rm B}^0$ in the TMA self-energies is replaced by
\begin{equation}
{\tilde G}_{\rm B}({\bm p},i\omega_n^{\rm B})=
{1 \over i\omega_n^{\rm B}-(\varepsilon_{\bm p}-\mu_{\rm B})-\Sigma_{\rm B}(0,0)},
\label{eq.10}
\end{equation}
so as to recover the required gapless Bose excitations at $T_{\rm BEC}$. In iTMA, one can carry out the Matsubara-frequency summation in the pair-correlation function in Eq. (\ref{eq.7}), giving
\begin{equation}
\Pi_{\rm BF}({\bm q},i\omega_n^{\rm F})
=-\sum_{\bm k}
{1-f(\xi_{{\bm k}+{\bm q}/2}^{\rm F})
+n_{\rm B}({\tilde \xi}_{-{\bm k}+{\bm q}/2}^{\rm B})
\over
i\omega_n^{\rm F}
-\xi_{{\bm k}+{\bm q}/2}^{\rm F}-{\tilde \xi}_{-{\bm k}+{\bm q}/2}^{\rm B}},
\label{eq.10b}
\end{equation}
where ${\tilde \xi}_{\bm p}^{\rm B}=\varepsilon_{\bm p}-{\tilde \mu}_{\rm B}$, with ${\tilde \mu}_{\rm B}=\mu_{\rm B}-\Sigma_{\rm B}(0,0)$. $f(\omega)$ and $n_{\rm B}(\omega)$ are the Fermi and the Bose distribution function, respectively.
\par
In both TMA and iTMA, one solves Eq. (\ref{eq.9}), together with the number equations,
\begin{equation}
N_{\rm B}=-T\sum_{{\bm p},\omega_n^{\rm B}}
G_{\rm B}({\bm p},i\omega_n^{\rm B}),
\label{eq.11}
\end{equation}
\begin{equation}
N_{\rm F}=T\sum_{{\bm p},\omega_n^{\rm F}}
G_{\rm F}({\bm p},i\omega_n^{\rm F}),
\label{eq.12}
\end{equation}
to self-consistently determine $T_{\rm BEC}$, $\mu_{\rm B}(T_{\rm BEC})$, and $\mu_{\rm F}(T_{\rm BEC})$. Thus, the different results between these two strong-coupling theories shown in Fig. \ref{fig1} purely come from the recovery of the Hugenholtz-Pines condition in the Bose Green's function in the iTMA self-energies. Of course, one can further improve iTMA by replacing all the Green's functions in the self-energies by the dressed ones, which remains as our future problem.
\par
Above $T_{\rm BEC}$, we only deal with the number equations (\ref{eq.11}) and (\ref{eq.12}), to obtain $\mu_{\rm B}(T)$ and $\mu_{\rm F}(T)$ shown in Fig. \ref{fig3}. Using these data, we calculate the single-particle spectral weight $A_{{\rm s}={\rm B,F}}({\bm p},\omega)$, as well as the single-particle density of states $\rho_{{\rm s}={\rm B,F}}(\omega)$, from the analytic continued Green's function as, respectively,
\begin{equation}
A_{\rm s}({\bm p},\omega)=
-{1 \over \pi}{\rm Im}
\left[
G_{\rm s}({\bm p},i\omega_n^{\rm s}\to\omega_+)
\right],
\label{eq.13}
\end{equation}
\begin{equation}
\rho_{\rm s}(\omega)=\sum_{\bm p}A_{\rm s}({\bm p},\omega),
\label{eq.14}
\end{equation}
where $\omega_+=\omega+i\delta$, with $\delta$ being an infinitesimally small positive number.
\par
\begin{figure}
\begin{center}
\includegraphics[width=9cm]{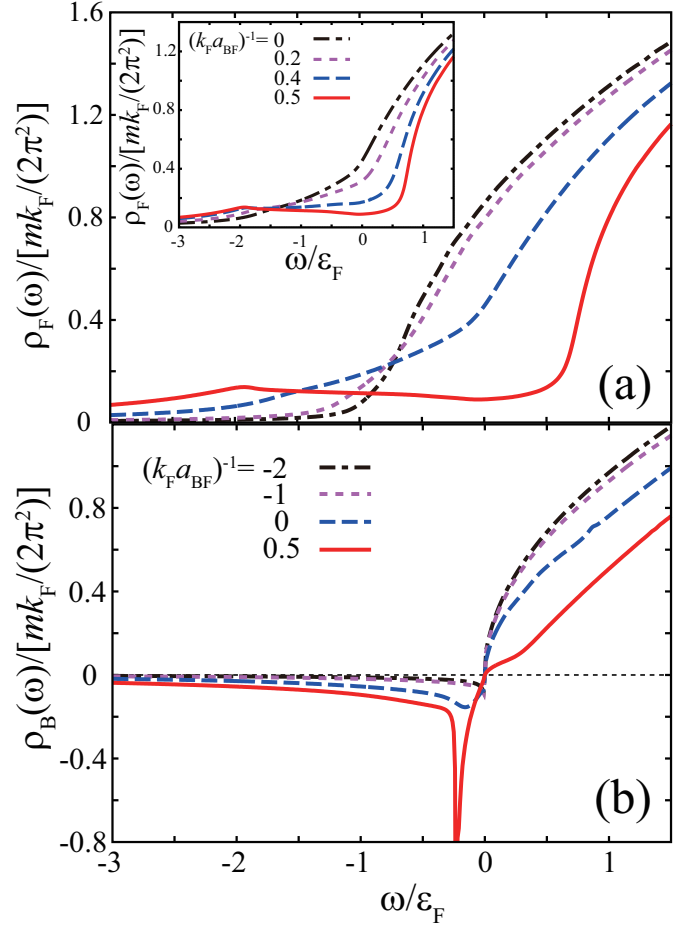}
\caption{(Color online) Calculated single-particle density of states in a Bose-Fermi mixture at $T_{\rm BEC}$. (a) Fermion component $\rho_{\rm F}(\omega)$. (b) Boson component $\rho_{\rm B}(\omega)$. Except for the inset, the interaction strengths used in this figure equal those at A$\sim$D in Fig. \ref{fig1}(a).}
\label{fig4}
\end{center}
\end{figure}
\par
\section{Single-particle properties of a Bose-Fermi mixture}
\par
Figure \ref{fig4}(a) shows the Fermi density of states $\rho_{\rm F}(\omega)$ in a Bose-Fermi mixture at $T_{\rm BEC}$. With increasing the interaction strength, the density of states $\rho_{\rm F}(\omega)$ around $\omega=0$ is found to gradually decrease. However, although TMA is known to give the pseudo-gapped density of states in a unitary Fermi gas near $T_{\rm c}$\cite{Tsuchiya2009}, such a dip structure does not appears in the case of a Bose-Fermi mixture at the unitarity ($(k_{\rm F}a_{\rm BF})^{-1}=0$). 
\par
In Fig. \ref{fig4}(a), a shallow dip structure is seen around $\omega=0$ when $(k_{\rm F}a_{\rm BF})^{-1}=0.5~(>0)$. In this regard, we note that, deep inside the strong-coupling regime (($k_{\rm F}a_{\rm BF})^{-1}\gg 1$), the system is reduced to an ideal Fermi gas of $N~(=N_{\rm F}=N_{\rm B})$ two-body bound molecules with the binding energy $E_{\rm bind}=-1/(ma_{\rm BF}^2)$. Since these fermions form a Fermi surface with the Fermi energy $E_{\rm F}=\varepsilon_{\rm F}/2$ (where $\varepsilon_{\rm F}$ is the Fermi energy of $N$ Fermi atoms), the dissociation energy $\omega=|E_{\rm bind}|-E_{\rm F}~(>0)$ is necessary to produce a Fermi and a Bose atom. Thus, the shallow dip structure seen in Fig. \ref{fig4}(a) when $(k_{\rm F}a_{\rm BF})^{-1}=0.5$ is considered to reflect that a Bose-Fermi mixture gradually changes to a gas of Fermi molecules in the strong-coupling side ($(k_{\rm F}a_{\rm BF})^{-1}>0$). 
\par
\begin{figure}
\begin{center}
\includegraphics[width=12cm]{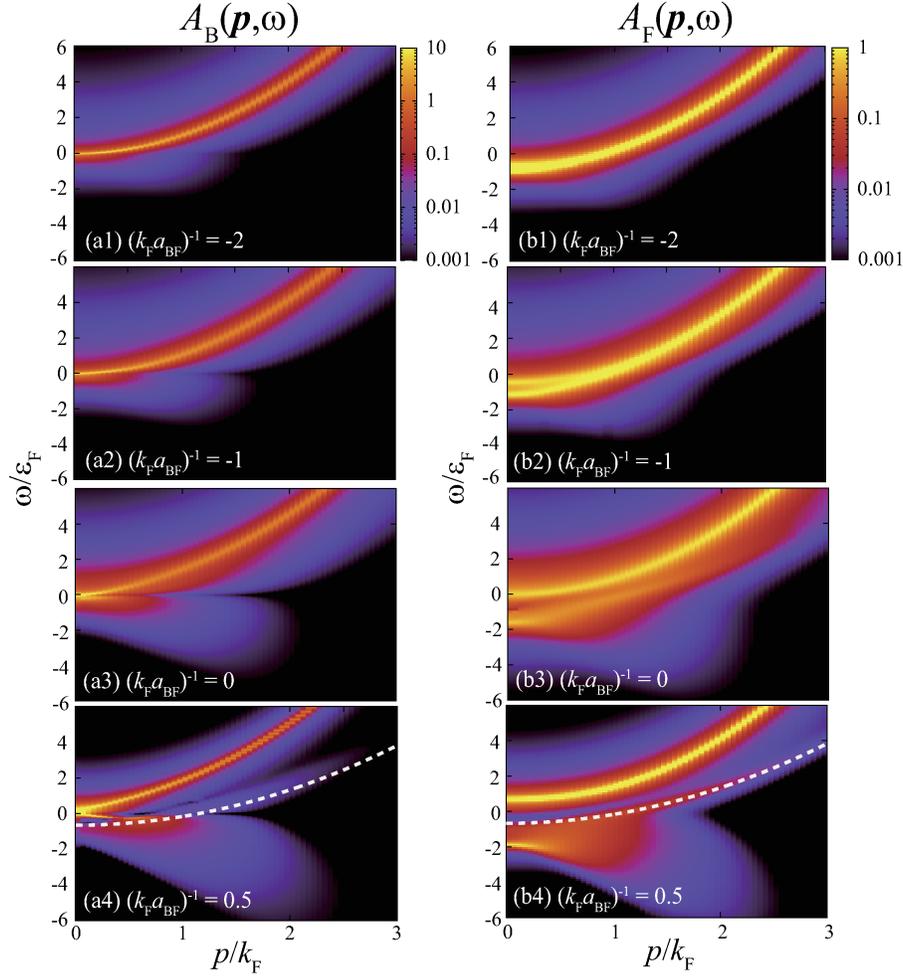}
\caption{(Color online) Calculated intensity of single-particle spectral weight at $T_{\rm BEC}$. (a1)-(a4) $A_{\rm B}({\bm p},\omega)$. (b1)-(b4) $A_{\rm F}({\bm p},\omega)$. Because the Bose spectral weight $A_{\rm B}({\bm p},\omega)$ is negative when $\omega<0$, we plot ${\rm sgn}(\omega)\times A_{\rm B}({\bm p},\omega)$ in panels (a1)-(a4). The interaction strengths used in this figure equal those at A$\sim$D in Fig. \ref{fig1}(a). In panels (a4) and (b4), the dashed lines show the molecular dispersion $\omega={\bm p}^2/(2M)-\mu_{\rm CF}$, where $M=2m$ is a molecular mass. The Fermi molecular chemical potential $\mu_{\rm CF}$ is determined as the peak energy of the spectrum, $-(1/\pi){\rm Im}[\Gamma_{\rm BF}({\bm p}=0,i\omega_n^{\rm F}\rightarrow\omega+i\delta)]$. The spectral intensity is normalized by the atomic Fermi energy $\varepsilon_{\rm F}$. This normalization is also used in Figs. \ref{fig6} and \ref{fig7}.}
\label{fig5}
\end{center}
\end{figure}
\par
Figure \ref{fig4}(b) shows the density of states $\rho_{\rm B}(\omega)$ of Bose atoms at $T_{\rm BEC}$. Starting from the weak-coupling regime, one sees that $\rho_{\rm B}(\omega)$ in the region $0\lesssim \omega/\varepsilon_{\rm F}\lesssim 1$ is gradually suppressed, as one passes through the unitarity limit ($(k_{\rm F}a_{\rm BF})^{-1}=0$). This also indicates that the system approaches a molecular Fermi gas in the strong-coupling regime, where finite dissociation energy is necessary to excite a Bose atom. 
\par
In addition to this, Fig. \ref{fig4}(b) also shows the negative Bose density of states $\rho_{\rm B}(\omega)<0$ in the negative energy region ($\omega<0$), which becomes more remarkable, as one increases the interaction strength. In particular, $\rho_{\rm B}(\omega)$ exhibits a negative peak around $\omega/\varepsilon_{\rm F}=-0.25$, when $(k_{\rm F}a_{\rm BF})^{-1}=0.5$. Since the density of states $\rho_{\rm B}(\omega)$ is given by the momentum-summation of the Bose spectral weight $A_{\rm B}({\bm p},\omega)$ (see Eq. (\ref{eq.14})), this peak structure implies the existence of strong Bose spectral intensity there. Indeed, Figs. \ref{fig5}(a1)-(a4) shows the growth of the spectral structure in the negative energy region of $A_{\rm B}({\bm p},\omega)$ with increasing the interaction strength. Apart from details, this phenomenon is also seen in the fermion component, as shown in Figs. \ref{fig5}(b1)-(b4). 
\par
To understand background physics of the spectral structures seen in Figs.\ref{fig5}(a4) and (b4), it is convenient to approximately treat the boson-fermion scattering matrix $\Gamma_{\rm BF}({\bm q},i\omega_n^{\rm F})$ in Eq. (\ref{eq.6}) as,
\begin{equation}
\Gamma_{\rm BF}({\bm q},i\omega_n^{\rm F})\simeq 
{\alpha_{\rm BF} \over i\omega_n^{\rm F}-\xi_{\bm q}^{\rm CF}}.
\label{eq.15}
\end{equation}
Here, $\xi_{\bm q}^{\rm CF}={\bm q}^2/(2M)-\mu_{\rm CF}$ is the kinetic energy of a composite Fermi molecule, measured from the molecular chemical potential $\mu_{\rm CF}$ (where $M=2m$ is a molecular mass). Strictly speaking, Eq. (\ref{eq.15}) is justified in the strong-coupling limit where the molecular dissociation no longer occurs (where $\alpha_{\rm BF}$ is given as $\alpha_{\rm BF}=8\pi/(m^2a_{\rm BF})>0$). However, this simple approximation is still helpful to grasp strong-coupling effects associated with hetero-pairing fluctuations in the unitary regime. Substituting Eq. (\ref{eq.15}) into Eqs. (\ref{eq.4}) and (\ref{eq.5}), we obtain, after replacing $G_{\rm B}^0$ by ${\tilde G}_{\rm B}$ in Eq. (\ref{eq.10})  and carrying out the summation over the fermion Matsubara frequencies,
\begin{equation}
\Sigma_{\rm B}({\bm p},i\omega_n^{\rm B})=\alpha_{\rm BF}
\sum_{\bm q}
\left[
{f(\xi_{\bm q}^{\rm F})
\over
i\omega_n^{\rm B}-\xi_{{\bm p}-{\bm q}}^{\rm CF}+\xi_{\bm q}^{\rm F}
}
-
{f(\xi_{\bm q}^{\rm CF})
\over
i\omega_n^{\rm B}+\xi_{{\bm p}-{\bm q}}^{\rm F}-\xi_{\bm q}^{\rm CF}
}
\right],
\label{eq.16}
\end{equation}
\begin{equation}
\Sigma_{\rm F}({\bm p},i\omega_n^{\rm F})=\alpha_{\rm BF}
\sum_{\bm q}
\left[
{n_{\rm B}({\tilde \xi}_{\bm q}^{\rm B})
\over
i\omega_n^{\rm F}-\xi_{{\bm p}-{\bm q}}^{\rm CF}+{\tilde \xi}_{\bm q}^{\rm B}
}
+
{f(\xi_{\bm q}^{\rm CF})
\over
i\omega_n^{\rm F}+{\tilde \xi}_{{\bm p}-{\bm q}}^{\rm B}-\xi_{\bm q}^{\rm CF}
}
\right].
\label{eq.17}
\end{equation}
Noting that (1) the Bose distribution function $n_{\rm B}({\tilde \xi}_{\bm q}^{\rm B})$ diverges at $T_{\rm BEC}$, and (2) $\mu_{\rm F}(T_{\rm BEC})/\varepsilon_{\rm F}\ll 1$ when $(k_{\rm F}a_{\rm BF})^{-1}=0.5$ (see Fig. \ref{fig1}(b)), we approximately set ${\bm q}=0$ in the denominator of the first term in each Eqs. (\ref{eq.16}) and (\ref{eq.17}). For the second term in each of these equations, we approximate the momentum ${\bm q}$ in the denominator to an ``effective Fermi momentum" ${\tilde {\bm p}}_{\rm F}^{\rm CF}$ of molecular fermions, for simplicity (where $|{\tilde {\bm p}}_{\rm F}^{\rm CF}|=\sqrt{2M\mu_{\rm CF}}$). Then, the analytic-continued Bose and Fermi Green's functions in iTMA are given by, respectively,
\begin{equation}
G_{\rm B}({\bm p},\omega_+)
\simeq
{1 \over
\displaystyle
\omega_+-\xi_{\bm p}^{\rm B}
-
{\lambda_{\rm F} \over \omega_+-\xi_{\tilde {\bm p}}^{\rm CF}}
+
\left\langle
{\lambda_{\rm CF} \over 
\omega_++\xi_{{\bm p}-{\tilde {\bm p}}_{\rm F}^{\rm CF}}^{\rm F}}
\right\rangle},
\label{eq.18}
\end{equation}
\begin{equation}
G_{\rm F}({\bm p},\omega_+)
\simeq
{1 \over
\displaystyle
\omega_+-\xi_{\bm p}^{\rm F}
-
{\lambda_{\rm B} \over \omega_+-\xi_{\bm p}^{\rm CF}}
-
\left\langle
{\lambda_{\rm CF} \over 
\omega_++{\tilde \xi}_{{\bm p}-{\tilde {\bm p}}_{\rm F}^{\rm CF}}^{\rm B}}
\right\rangle},
\label{eq.19}
\end{equation}
where $\lambda_{\rm B}=\alpha_{\rm BF} N_{\rm B}^0$, $\lambda_{\rm F}=\alpha_{\rm BF} N_{\rm F}^0$, and $\lambda_{\rm CF}=\alpha_{\rm BF} N_{\rm CF}^0$, with $N_{\rm B}^0=\sum_{\bm p}n_{\rm B}({\tilde \xi}_{\bm p}^{\rm B})$, $N_{\rm F}^0=\sum_{\bm p}f(\xi_{\bm p}^{\rm F})$, and $N_{\rm CF}^0=\sum_{\bm p}f(\xi_{\bm p}^{\rm CF})$. In Eqs. (\ref{eq.18}) and (\ref{eq.19}), $\langle\cdot\cdot\cdot\rangle$ means the average over the direction of ${\tilde {\bm p}}_{\rm F}^{\rm CF}$. Equation (\ref{eq.18}) and (\ref{eq.19}) clearly indicate that hetero-pairing fluctuations cause a coupling phenomenon among Fermi atomic excitations ($\xi_{\bm p}^{\rm F}$), Bose atomic excitations (${\tilde \xi}_{\bm p}^{\rm B}$), and Fermi molecular excitations ($\xi_{\bm p}^{\rm CF}$), with the coupling constants $\lambda_{{\rm s}={\rm B,F,CF}}$. This explains the appearance of the spectral peak along the molecular dispersion, $\omega=\xi_{\bm p}^{\rm CF}={\bm p}/(2M)-\mu_{\rm CF}$, in Figs. \ref{fig5}(a4) and (b4) (dotted lines).
\par
In addition to this molecular contribution, Eq. (\ref{eq.18}) indicates the existence of the contribution of fermionic hole excitations to the Bose spectral weight $A_{\rm B}({\bm p},\omega)$. Because of the average over the direction of ${\tilde {\bm p}}_{\rm F}^{\rm CF}$, this contribution gives a broad spectral structure in $A_{\rm B}({\bm p},\omega)$ around 
\begin{equation}
-{(p+{\tilde p}_{\rm F}^{\rm CF})^2 \over 2m}
\le \omega\le 
-{(p-{\tilde p}_{\rm F}^{\rm CF})^2 \over 2m},
\label{eq.20}
\end{equation}
where we have ignored $\mu_{\rm F}~(\ll\varepsilon_{\rm F})$, because we are considering the strong-coupling case in Fig. \ref{fig5}(a4). Indeed, such a broad spectral structure is seen in the negative energy region in Fig. \ref{fig5}(a4). This spectral weight, as well as the molecular contribution around ${\bm p}=0$ (where $\xi_{\bm p}^{\rm CF}<0$), gives the negative Bose density of states with the negative peak structure seen in Fig. \ref{fig4}(b) when $(k_{\rm F}a_{\rm BF})^{-1}=0.5$.
\par
\begin{figure}
\begin{center}
\includegraphics[width=9cm]{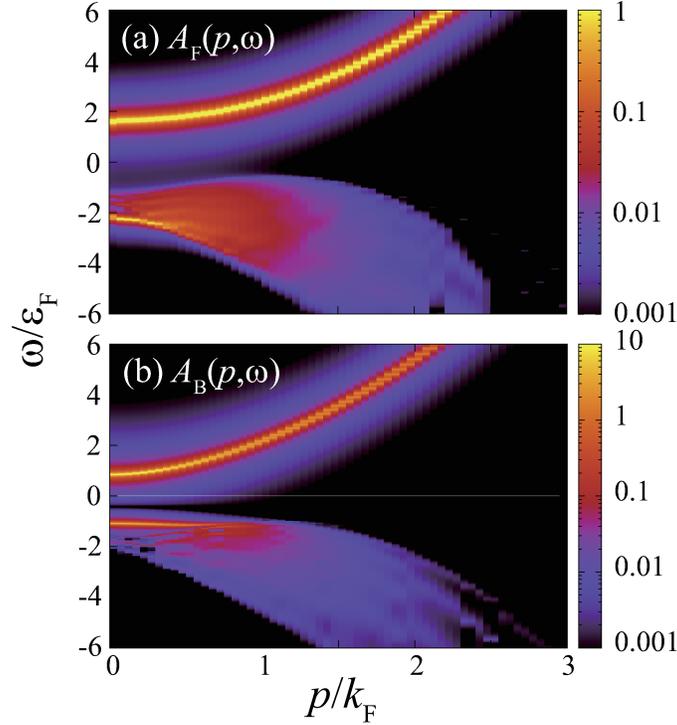}
\caption{(Color online) Intensity of singe-particle spectral weight in the strong-coupling regime, when $(k_{\rm F}a_{\rm BF})^{-1}=1$ (``E" in Fig. \ref{fig1}(a)). We take $T/T_{\rm F}=0.0036$. (a) Bose component $A_{\rm B}({\bm p},\omega)$. (b) Fermi component $A_{\rm F}({\bm p},\omega)$.}
\label{fig6}
\end{center}
\end{figure}
\par
As in the Bose case, the last term of the denominator in the Fermi Green's function in Eq. (\ref{eq.19}) also gives a broad hole-type spectral structure in the negative energy region of the Fermi spectral weight $A_{\rm F}({\bm p},\omega)$. At a glance, this spectral intensity and the spectral structure in the positive energy region coming from the ordinary Fermi particle dispersion $\omega=\xi_{\bm p}^{\rm F}$ may give a pseudogap structure in the density of states $\rho_{\rm F}(\omega)$ around $\omega=0$. However, the above-mentioned Fermi molecular dispersion, which passes through $\omega=0$, increases $\rho_{\rm F}(\omega\sim 0)$, so that $\rho_{\rm F}(\omega)$ actually does not exhibit such a dip structure in the unitariry limit. Because of the same reason, although a dip structure appears in $\rho_{\rm F}(\omega\sim 0)$ when ($k_{\rm F}a_{\rm BF})^{-1}=0.5$, it is very shallow, as shown in Fig. \ref{fig4}(a).
\par
Figure \ref{fig6} shows the spectral weight $A_{{\rm s}={\rm B,F}}({\bm p},\omega)$ at ``E" in Fig.\ref{fig1}(a). In this figure, the spectral peak along the molecular $\omega=\xi_{\bm p}^{\rm CF}$ is invisible, in contrast to Fig. \ref{fig5}(a4) and (b4). As a result, we see a clear gap-like structure around $\omega=0$ in each Bose and Fermi component in Fig. \ref{fig6}.  
\par
To explain the reason for the disappearance of the molecular dispersion in Fig. \ref{fig6}, we recall that the BEC phase transition vanishes in the strong-coupling regime when $(k_{\rm F}a_{\rm BF})^{-1}\ge 0.56$ (see Fig. \ref{fig1}(a)). In addition, when $(k_{\rm F}a_{\rm BF})^{-1}=1$, the Fermi chemical potential $\mu_{\rm F}$, as well as the effective Bose chemical potential ${\tilde \mu}_{\rm B}=\mu_{\rm B}-\Sigma_{\rm B}(0,0)$, are negative (see Fig. \ref{fig3}). Thus, in the case of Fig. \ref{fig6} ($(k_{\rm F}a_{\rm BF})^{-1}=1$ and $T/T_{\rm F}=0.0036\ll 1$), the coupling constant $\lambda_{\rm F}=\alpha_{\rm BF}\sum_{\bm p}f(\varepsilon_{\bm p}-\mu_{\rm F})$ in Eq. (\ref{eq.18}), as well as $\lambda_{\rm B}=\alpha\sum_{\bm p}n_{\rm B}(\varepsilon_{\bm p}-{\tilde \mu}_{\rm B})$ in Eq. (\ref{eq.19}), vanish in the limit $T\to 0$, which immediately explains the vanishing molecular spectral peak in Fig. \ref{fig6}. On the other hand,  because the molecular fermions form a Fermi surface in the strong-coupling regime, the molecular chemical potential $\mu_{\rm CF}$ is positive in Fig. \ref{fig6}. Thus, the coupling constant $\lambda_{\rm CF}=\alpha_{\rm BF}f(\xi_{\bm p}^{\rm CF})$ does {\it not} vanish, leading to the non-vanishing spectral structure in the negative energy region of the spectral weight there.
\par
In Fig. \ref{fig6}, when we estimate the (pseudo)gap energy $\omega_{\rm gap}^{\rm F}$ in the Fermi spectral weight $A_{\rm F}({\bm p},\omega)$, as well as the gap energy $\omega_{\rm gap}^{\rm B}$ in the Bose spectral weight $A_{\rm B}({\bm p},\omega)$, from the peak-to-peak energy at ${\bm p}=0$ in panel (a) and (b), respectively, their magnitudes are found to be different as
\begin{eqnarray}
\left\{
\begin{array}{l}
\omega_{\rm gap}^{\rm B}=1.9\varepsilon_{\rm F},\\
\omega_{\rm gap}^{\rm F}=3.7\varepsilon_{\rm F},
\end{array}
\right.
\label{eq.19b}
\end{eqnarray}
in spite of the fact that both the energy gaps are associated with the dissociation of a Bose-Fermi molecule. 
\par
To understand the origin of this difference, the approximate Green's functions in Eqs. (\ref{eq.18}) and (\ref{eq.19}) are also helpful. Setting $\lambda_{\rm F}=\lambda_{\rm B}=0$ in Eq. (\ref{eq.18}), we obtain the two bosonic eigen-energies $\omega_\pm^{\rm B}$ at ${\bm p}=0$ as,
\begin{equation}
\omega_{\pm}^{\rm B}=
{1 \over 2}
\left[
[|\mu_{\rm B}|-|\mu_{\rm F}|-\varepsilon_{{\tilde {\bm p}}_{\rm F}^{\rm CF}}]
\pm
\sqrt{(|\mu_{\rm B}|+|\mu_{\rm F}|+\varepsilon_{{\tilde {\bm p}}_{\rm F}^{\rm CF}})^2-4\lambda_{\rm CF}}
\right].
\label{eq.21}
\end{equation}
In the same manner, Eq.(\ref{eq.19}) gives the two fermionic eigen-energies $\omega_\pm^{\rm F}$ at ${\bm p}=0$,
\begin{equation}
\omega_{\pm}^{\rm F}=
{1 \over 2}
\left[
[|\mu_{\rm F}|-|{\tilde \mu}_{\rm B}|-\varepsilon_{{\tilde {\bm p}}_{\rm F}^{\rm CF}}]
\pm
\sqrt{(|{\tilde \mu}_{\rm B}|+|\mu_{\rm F}|+\varepsilon_{{\tilde {\bm p}}_{\rm F}^{\rm CF}})^2+4\lambda_{\rm CF}}
\right].
\label{eq.21b}
\end{equation}
Using these results, one has
\begin{eqnarray}
\left\{
\begin{array}{l}
\omega_{\rm gap}^{\rm B}=\sqrt{(|\mu_{\rm B}|+|\mu_{\rm F}|+\varepsilon_{{\tilde {\bm p}}_{\rm F}^{\rm CF}})^2-4\lambda_{\rm CF}},
\\
\omega_{\rm gap}^{\rm F}=\sqrt{(|{\tilde \mu}_{\rm B}|+|\mu_{\rm F}|+\varepsilon_{{\tilde {\bm p}}_{\rm F}^{\rm CF}})^2+4\lambda_{\rm CF}}.
\end{array}
\right.
\label{eq.22}
\end{eqnarray}
To estimate Eqs. (\ref{eq.22}) in the case of Fig. \ref{fig6}, we employ the strong-coupling expression $\alpha_{\rm BF}=8\pi/(m^2a_{\rm BF})$ in $\lambda_{\rm CF}=\alpha_{\rm BF}N_{\rm CF}^0$, and assume that all the atoms form Fermi molecules ($N_{\rm CF}^0\simeq N_{\rm F})$. Substituting the iTMA values, $\mu_{\rm F}=-0.92\varepsilon_{\rm F}$, $\mu_{\rm B}=-1.59\varepsilon_{\rm F}$, and ${\tilde \mu}_{\rm B}=-0.51\varepsilon_{\rm F}$ into Eq. (\ref{eq.22}), we obtain 
\begin{eqnarray}
\left\{
\begin{array}{l}
\omega_{\rm gap}^{\rm B}=2.35\varepsilon_{\rm F},\\
\omega_{\rm gap}^{\rm F}=3.56\varepsilon_{\rm F}.
\end{array}
\right.
\label{eq.19c}
\end{eqnarray}
This rough estimation gives comparable gap sizes to Eq. (\ref{eq.19b}). 
\par
In the extreme strong-coupling regime ($(k_{\rm F}a_{\rm BF})^{-1}\gg 1$), the value of $\mu_{\rm F}+\mu_{\rm B}$ would approach the binding energy $E_{\rm bind}=1/(ma_{\rm BF}^2)~(\gg \varepsilon_{\rm F})$ of a tightly bound molecule. In this limit, one may safely ignore other terms in Eq. (\ref{eq.22}), giving $\omega_{\rm gap}^{\rm B}=\omega_{\rm gap}^{\rm F}=E_{\rm bind}$, as expected. 
\par
To clarify the character of strong-coupling effects in a Bose-Fermi mixture, it is helpful to compare our results with those in the case of a two-component Fermi gas. In the latter, the TMA self-energy $\Sigma_{\rm F}({\bm p},i\omega_n^{\rm F})$ in the Fermi single-particle Green's function has the form,
\begin{equation}
\Sigma_{\rm F}({\bm p},i\omega_n)=T\sum_{{\bm q},\omega_{n'}^{\rm B}}
\Gamma_{\rm FF}({\bm q},i\omega_{n'}^{\rm B})
G_{\rm F}^0({\bm q}-{\bm p},i\omega_{n'}^{\rm B}-i\omega_n^{\rm F}).
\label{eq.23}
\end{equation}
Here, $\Gamma_{\rm FF}({\bm q},i\omega_n^{\rm B})$ is the TMA fermion-fermion scattering matrix, describing fluctuations in the Cooper-channel (For the detailed expression, see, for example, Ref. \cite{Tsuchiya2009}). As in the case of Bose-Fermi mixture, for simplicity, we approximate $\Gamma_{\rm FF}({\bm q},i\omega_n^{\rm B})$ to the Cooper-pair propagator,
\begin{equation}
\Gamma_{\rm FF}({\bm q},i\omega_n^{\rm B})=
{\alpha_{\rm FF} \over i\omega_n^{\rm B}-\xi_{\bm q}^{\rm CB}},
\label{eq.24}
\end{equation}
where $\xi_{\bm q}^{\rm CB}={\bm q}^2/(2M)-\mu_{\rm CB}$ is the kinetic energy of a Cooper pair with the molecular mass $M=2m$ and the chemical potential $\mu_{\rm CF}$. The factor $\alpha_{\rm FF}$ is reduced to $\alpha_{\rm FF}=8\pi/(m^2a_{\rm FF})$ in the strong-coupling BEC limit\cite{Strinati}, where $a_{\rm FF}$ is the $s$-wave scattering length for a contact-type interaction between Fermi atoms. Substituting Eq. (\ref{eq.24}) into (\ref{eq.23}), we have
\begin{equation}
\Sigma_{\rm FF}({\bm p},i\omega_n^{\rm F})=
\alpha_{\rm FF}
\sum_{\bm q}
\left[
{n_{\rm B}(\xi_{\bm q}^{\rm CB})
\over
i\omega_n^{\rm F}+\xi_{{\bm p}-{\bm q}}^{\rm F}-\xi_{\bm q}^{\rm CB}
}
+
{f(\xi_{\bm q}^{\rm F})
\over
i\omega_n^{\rm F}+\xi_{\bm q}^{\rm F}-\xi_{{\bm p}-{\bm q}}^{\rm CB}
}
\right].
\label{eq.25}
\end{equation} 
At the superfluid phase transition temperature $T_{\rm c}$, the Cooper-pair chemical potential $\mu_{\rm CB}$ vanishes, according to the Thouless criterion\cite{Thouless}. Noting this, we approximately set ${\bm q}=0$ in the denominator of the first term in Eq. (\ref{eq.25}). For the denominator of the second term in Eq. (\ref{eq.25}), we approximate ${\bm q}$ to the ``effective Fermi momentum" ${\tilde {\bm p}}_{\rm F}$ of Fermi atoms, where $|{\tilde {\bm p}}_{\rm F}|=\sqrt{2m\mu_{\rm F}}$. (We consider the unitary regime, where the Fermi chemical potential is still positive.) The TMA single-particle Fermi Green's function corresponding to Eq. (\ref{eq.22}) is then approximated to
\begin{equation}
G_{\rm F}({\bm p},\omega_+)
\simeq
{1 \over
\displaystyle
\omega_+-\xi_{\bm p}^{\rm F}
-
{\lambda_{\rm CB} \over \omega_++\xi_{\bm p}^{\rm F}}
-
\left\langle
{\lambda_{\rm F} \over 
\omega_+-\xi_{{\bm p}-{\tilde {\bm p}}_{\rm F}}^{\rm CB}}
\right\rangle}.
\label{eq.26}
\end{equation}
Equation (\ref{eq.26}) shows that bosonic fluctuations in the Cooper channel couple Fermi atomic dispersion ($\omega=\xi_{\bm p}^{\rm F}$) with the hole dispersion ($\omega=-\xi_{\bm p}^{\rm F}$) with the coupling constant $\lambda_{\rm CB}=\alpha_{\rm FF}\sum_{\bm q}n_{\rm B}(\xi_{\bm q}^{\rm CB})$. When we only retain this effect, Eq. (\ref{eq.26}) gives the BCS-like gapped single-particle dispersions, $\omega_\pm =\pm\sqrt{\xi_{\bm p}^2+\lambda_{\rm CB}}$, where $\lambda_{\rm CB}~(>0)$ plays a similar role to the square $\Delta^2$ of the BCS superfluid order parameter. ($\lambda_{\rm CB}$ is sometimes referred to as the pseudogap parameter\cite{Levin}). In a Bose-Fermi mixture, the corresponding coupling phenomenon is brought about by Bose atomic excitations, which is characterized by $\lambda_{\rm B}$ in Eq. (\ref{eq.19}). However, what is coupled with the Fermi atomic dispersion by $\lambda_{\rm B}$ is the Fermi molecular dispersion ($\omega=\xi_{\bm p}^{\rm CF}={\bm p}^2/(2M)-\mu_{\rm CF}$), passing through $\omega=0$. Thus, while the particle-hole coupling in a two-component Fermi gas suppresses the single-particle density of states $\rho_{\rm F}(\omega\sim 0)$ (pseudogap phenomenon), the Fermi-molecule coupling in a Bose Fermi mixture enhances $\rho_{\rm F}(\omega\sim 0)$.
\par
The last term in the denominator in Eq. (\ref{eq.26}) corresponds to that in Eq. (\ref{eq.19}). However, while the latter gives a broad spectral structure in the {\it negative} energy region of the spectral weight (see Fig. \ref{fig5}(a4)), the former produces a spectral structure in the {\it positive} energy region around
\begin{equation}
{(p-{\tilde p}_{\rm F})^2 \over 2M}\le\omega\le {(p+{\tilde p}_{\rm F})^2 \over 2M}~~(T=T_{\rm c}).
\label{eq.27}
\end{equation}
This broad spectral structure has not been frequently discussed in cold Fermi gas physics, because the spectral weight in the positive energy region is usually dominated by the strong peak intensity along the particle dispersion ($\omega=\xi_{\bm p}^{\rm F}$). However, it has been pointed out\cite{Inotani} that the particle dispersion becomes broad, when it is in the region in Eq. (\ref{eq.27}), because of the coupling phenomenon described by $\lambda_{\rm F}$ in Eq. (\ref{eq.26}).
\par
\begin{figure}
\begin{center}
\includegraphics[width=12cm]{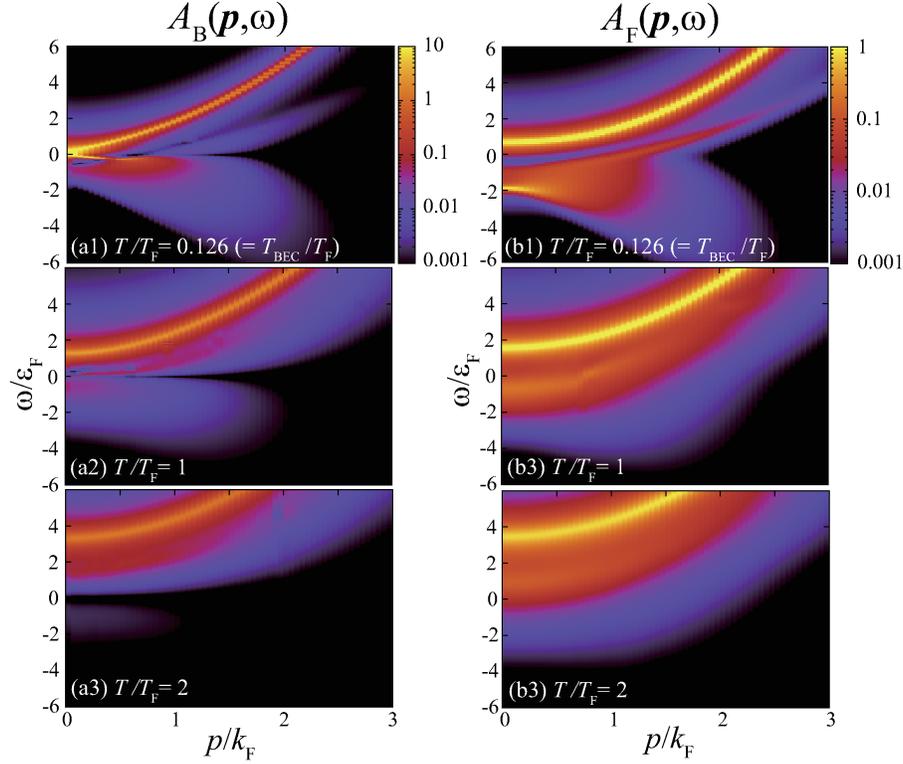}
\caption{(Color online) Single-particle spectral weight above $T_{\rm BEC}$. We take $(k_{\rm F}a_{\rm BF})^{-1}=0.5$. (a1)-(a3) $A_{\rm B}({\bm p},\omega)$. (b1)-(b3) $A_{\rm F}({\bm p},\omega)$.
}
\label{fig7}
\end{center}
\end{figure}
\par
\begin{figure}
\begin{center}
\includegraphics[width=9cm]{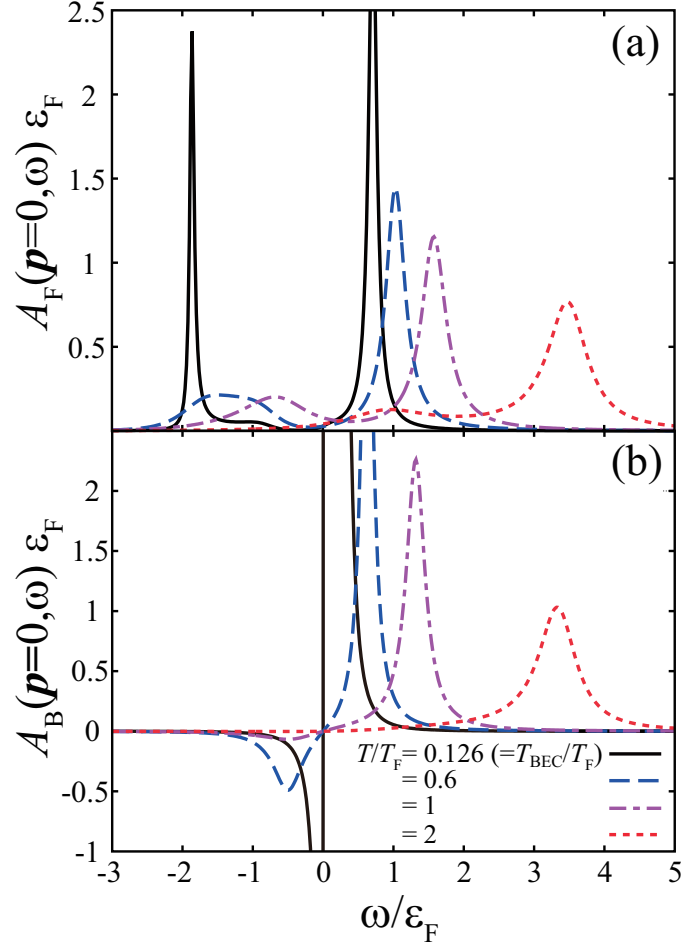}
\caption{(Color online) Single-particle spectral weight at ${\bm p}=0$, as a function of energy $\omega$. We take $(k_{\rm F}a_{\rm BF})^{-1}=0.5$. (a) $A_{\rm F}({\bm p}=0,\omega)$. (b) $A_{\rm B}({\bm p}=0,\omega)$.}
\label{fig8}
\end{center}
\end{figure}
\par
Finally, we examine the spectral weight $A_{{\rm s}={\rm B,F}}({\bm p},\omega)$ above $T_{\rm BEC}$. Figure \ref{fig7} shows that the strong-coupling phenomenon at $T_{\rm BEC}$ gradually disappears with increasing the temperature, to approach the expected ordinary spectral structure with a single peak along the free particle dispersion. From the viewpoint of Eqs. (\ref{eq.18}) and (\ref{eq.19}), this behavior can be understood as the result of the fact that all the coupling constants $\lambda_{{\rm s}={\rm B,F,CB}}$ become small because the chemical potentials $\mu_{\rm F}$, ${\tilde \mu}_{\rm B}$, and $\mu_{\rm CF}$, decrease with increasing the temperature. We briefly note that, although we only show the result at $(k_{\rm F}a_{\rm BF})^{-1}=0.5$ in Fig. \ref{fig7}, this tendency is also seen at other interaction strengths.
\par
One sees in Fig. \ref{fig7} that the spectral structure is still somehow different from the non-interacting case, even when $T\sim T_{\rm F}$. Indeed, when we plot the spectral weight at ${\bm p}=0$ as a function of the energy $\omega$, a clear double peak structure is seen even at $T/T_{\rm F}=1$ in $A_{\rm F}({\bm p}=0,\omega)$, and at $T/T_{\rm F}=0.6$ in $A_{\rm B}({\bm p}=0,\omega)$, as shown in Fig. \ref{fig8}. In this regard, we note that the photoemission-type experiment developed by JILA group\cite{Stewart2008,Gaebler2010,Sagi2015} can observe the spectral weight multiplied by the Fermi or Bose distribution function, depending on particle statistics. Thus, these anomalous spectral structures in the negative energy region may be observable even at relatively high temperatures by using this experimental technique. 
\par
\par
\section{Summary}
\par
To summarize, we have discussed strong-coupling properties of a Bose-Fermi mixture with a hetero-nuclear Feshbach resonance, Including hetero-pairing fluctuations within the framework of an improved $T$-matrix approximation\cite{Vijay}, we have calculated the single-particle density of states, as well as the single-particle spectral weight, in the normal state above $T_{\rm BEC}$.
\par
In the unitary regime at $T_{\rm BEC}$, we showed that hetero-pairing fluctuations near $T_{\rm BEC}$ cause a coupling phenomenon between Fermi atomic excitations and Fermi molecular excitations. This phenomenon is similar to the so-called particle-hole coupling known in the BCS-BEC crossover regime of a two-component Fermi gas, where fluctuations in the Cooper channel couple Fermi atomic excitations with hole excitations near $T_{\rm c}$. However, while the particle-hole coupling suppresses the density of states around $\omega=0$ in the latter Fermi system, the atom-molecule coupling in a Bose-Fermi mixture enhances the Fermi density of states around $\omega=0$, because the coupled molecular excitations passes through $\omega=0$.
\par
In addition to this atom-molecule coupling, we showed that hetero-pairing fluctuations also couples Fermi atomic excitations with Bose atomic excitations. This phenomenon affects the Fermi spectral weight $A_{\rm F}({\bm p},\omega)$ in the negative energy region. In the strong-coupling regime where the Bose-Einstein condensation no longer occurs, the atom-molecule coupling disappears at low temperatures, so that the spectral structure of $A_{\rm F}({\bm p},\omega)$ is dominated by Fermi atomic excitations in the positive energy region and Bose atomic excitations in the negative energy region.
\par
These coupling phenomena are also seen in the Bose density of states $\rho_{\rm B}(\omega)$, as well as the Bose spectral weight $A_{\rm B}({\bm p},\omega)$. In this case, Bose atomic excitations couple with Fermi molecular excitations, as well as Fermi hole excitations. These couplings lead to negative spectral intensity and negative Bose density of states in the negative energy region. 
\par
These strong-coupling phenomena gradually disappear with increasing the temperature above $T_{\rm BEC}$. However, we found that the spectral intensity in the negative energy region associated with the above-mentioned coupling phenomena remains up to relatively high temperatures. Thus, even when one cannot reach the BEC phase transition, the strong-coupling corrections to single-particle excitations may be observed by the photoemission-type experiment. It is an interesting future problem to clarify how hetero-pairing fluctuations affect photoemission spectra. 
\par
In this paper, we have ignored mass difference between Fermi and Bose atoms, as well as effects of a harmonic trap, for simplicity. Inclusion of these realistic situations also remains as our future problem. In addition, besides hetero-pairing fluctuations, a Bose-Fermi interaction $-U_{\rm BF}$ also induces intra-species interactions, mediated by Bose and Fermi density fluctuations, that are known to be important in considering the stability of a Bose-Fermi mixture\cite{Molmer,Miyakawa,Roth}. Thus, extension of the present work to include these induced interactions, as well as direct intra-species interactions, is a crucial future challenge. Since single-particle excitations are now observable in cold atom physics by using the photoemission-type technique, our results would be useful for the study of strong-coupling physics in a Bose-Fermi mixture from the viewpoint of single-particle properties of this system.
\par
\par
\begin{acknowledgements}
We thank H. Tajima, M. Matsumoto, P. van Wyk, and D. Kagamihara for discussions. This work was supported by KiPAS project in Keio University. DK thanks Japan International Co-operation Agency (JICA) and Keio Leading-edge Laboratory of Science and Technology (KLL) for supporting this research.  DI was supported by Grant-in-Aid for Young Scientists (B) (No. JP16K17773) from JSPS. RH was supported by Grant-in-Aid for JSPS fellows. YO was supported by Grand-in-Aid for Scientific Research from MEXT and JSPS in Japan (No. JP15K00178, No. JP15H00840, No. JP16K05503).
\end{acknowledgements}
\par


\begin{thebibliography}{99}
\bibitem{note1} When $(k_{\rm F}a_{\rm BF})^{-1}>0$, a Fermi and a Bose atom form a two-body bound molecule, with the binding energy $E_{\rm bind}=-1/(2m_{\rm r}a_{\rm BF}^2)$, where $m_{\rm r}$ is a reduced mass. In the strong-coupling regime where the Bose-Einstein condensation no longer occurs, the Fermi chemical potential $\mu_{\rm F}$ should become negative, to approach half the binding energy.
\end{thebibliography}
\end{document}